 \documentclass[final,1p,times]{elsarticle}

 \usepackage{graphicx}

\usepackage{amssymb}

 \usepackage{lineno}

\journal{Planetary and Space Science}

\begin{document}

\begin{frontmatter}





\title{The effects of the target material properties and layering on
  the crater chronology: the case of Raditladi and Rachmaninoff basins
  on Mercury}
%

%
\author[mar]{S. Marchi\corref{cor1}}
\ead{marchi@oca.eu}
\author[gia]{M. Massironi}
\author[cre]{G. Cremonese}
\author[mart]{E. Martellato}
\author[gia]{L. Giacomini}
\author[pro]{L. Prockter}
%
\cortext[cor1]{Corresponding author}
\address[mar]{Departement Cassiop\'{e}e,  Universite de Nice  - Sophia
  Antipolis, Observatoire  de la C\^{o}te d'Azur,  CNRS, Nice, France}
\address[gia]{Department    of  Geosciences, Padova University, Italy} 
\address[cre]{INAF-Padova, Italy}
\address[mart]{Center of Studies and Activities for Space, Padova University,    Italy}    
\address[pro]{Applied Physics Laboratory, Johns Hopkins University, USA}

\begin{abstract}

In this paper we present a crater age determination of several
terrains associated with the Raditladi and Rachmaninoff basins. These
basins were discovered during the first and third MESSENGER flybys of
Mercury, respectively. One of the most interesting features of both
basins is their relatively fresh appearance. The young age of both
basins is confirmed by our analysis on the basis of age determination
via crater chronology. The derived Rachmaninoff and Raditladi basin
model ages are about 3.6~Ga and 1.1~Ga, respectively.  Moreover, we also
constrain the age of the smooth plains within the basins' floors. This
analysis shows that Mercury had volcanic activity until recent time,
possibly to about 1~Ga or less.  We find that some of the crater
size-frequency distributions investigated suggest the presence of a
layered target.  Therefore, within this work we address the importance
of considering terrain parameters, as geo-mechanical properties and
layering, into the process of age determination.  We also comment on
the likelihood of the availability of impactors able to form basins
with the sizes of Rachmaninoff and Raditladi in relatively recent
times.

\end{abstract}


\end{frontmatter}



\section{Introduction}

During MESSENGER's third flyby of Mercury, a ~290-km-diameter
peak-ring (double-ring) impact basin, centered at 27.6$^\circ$~N,
57.6$^\circ$~E, was discovered and subsequently named Rachmaninoff. In
terms of size and morphology, the Rachmaninoff basin closely resembles
the 265-km-diameter Raditladi peak-ring basin, located at
27$^\circ$~N, 119$^\circ$~E west of the Caloris basin, that was
discovered during MESSENGER's first flyby \citep{pro09}. The
image-mosaic of Rachmaninoff and its ejecta has a spatial resolution
of 500~m/pixel and it is derived from images obtained by MESSENGER's
Mercury Dual Imaging System (MDIS) narrow-angle camera \citep{haw07},
while Raditladi basin and surrounding areas were imaged at
280~m/pixel. Both basins and surrounding areas were also imaged
  with a set of 11 filters of the MDIS wide-angle camera (WAC), whose
  wavelengths range from 430 to 1020~nm \citep{haw07}. These images
  were used to obtain color maps with a resolution of about 5~km/pixel
  and 2.4~km/pixel for Rachmaninoff and Raditladi, respectively.\\
The two basins appeared to be remarkably young because of the
small number of impact craters seen within their rims
\citep{str08,pro09,pro10}. For this reason it has been argued that
they were likely formed well after the end of the late heavy
bombardment of the inner Solar System at about 3.8~Ga \citep{str08}.
In particular, for Raditladi it has been pointed out that the basin
could be as young as 1~Ga or less \citep{str08,pro10}. \\ 
Interestingly, both basin floors are partially covered by smooth
plains.  In the case of Rachmaninoff, an inner floor filled with
spectrally distinct smooth plains has been observed and this, combined
with the small number of overimposed craters, implies a volcanic
origin \citep{pro10}.  The estimate of the temporal extent of the
volcanic activity and, in particular, the timing of the most recent
activity may represent a key element in our understanding of the
global thermal evolution of Mercury, and helps to constrain the
duration of the geologic activity on the planet in light of the new
data provided by MESSENGER. Moreover, Raditladi may be the
  youngest impact basin discovered on Mercury so far, and
  therefore it is important to understanding the recent impact history
  of the planet.\\
For all these reasons, the age determination of Rachmaninoff and
Raditladi basins and their geologically different terrains is of a
great interest. In this paper, we will present a revised Mercury
  crater chronology, and show how to take into account for the crustal
  properties of the target ($\S$~3). This chronology will be then
  applied to Rachmaninoff and Raditladi basins ($\S$~4).

\section{The Model Production Function chronology}

In this paper we date the Raditladi and Rachmaninoff basin units by
means of the the Model Production Function (MPF) chronology of Mercury
\citep{mar09,mas09}.  This chronology relies on the knowledge of the
impactor flux on Mercury and on the computed ratio of impactors
between Mercury and the Moon. The absolute age calibration is provided
by the Apollo sample radiometric ages.  The crater scaling law
  enables computation of the crater size-frequency distribution (SFD)
  using a combination of the impactor SFD and the inferred physical
  properties of the target. The computed crater SFD per unit surface
  and unit time is the so-called MPF.  The present model involves
  several improvements with respect to the model presented in
  \cite{mar09} and \cite{mas09}, thus it will be described in detail
  in the next sections.\\

\subsection{The impactor SFD and the crater scaling law}

In the following analysis, we use the present Near-Earth Object
  (NEO) population as the prime source of impactors.  This assumption is
  justified by the presumably young ages (i.e. low crater density) of
  the terrains studied in this paper.  In particular, we use the NEO
  SFD as modeled by \cite{bot00,bot02}.  This NEO SFD is in
  good agreement with the observed NEO population, fireballs and bolide
  events \citep[see][for further details]{mar09}.   \\
Concerning the crater scaling law, we adopted the  so-called Pi-scaling law 
in the formulation by \cite{hol07}. Unlike previous approaches, our
methodology explicitly takes into account the crustal properties of
the target. In fact, surfaces react differently to impact processes,
depending on the bulk density, strength and bulk structure of the
target material. These latter parameters are taken into account by the
scaling law, and are tabulated for several materials like cohesive
soils, hard-rock and porous materials \citep[e.g.][]{mel89,hol07}.
On a planetary body, terrain properties may vary from place to place
according to the local geological history and as a function of the
depth in the target crust.  Therefore, impacts of different sizes
taking place on a particular terrain may require different estimates
of the target properties.\\
The Pi-scaling law allows
computation of the transient crater diameter ($D_t$) as a function of
impact conditions and target properties, and reads:

\begin{equation}
D_t=kd\biggl[\frac{gd}{2v_{\perp}^2}\biggl(\frac{\rho}{\delta}\biggr)^\frac{2\nu}{
\mu}+\biggl(\frac{Y}{\rho v_{\perp}^2}\biggr)^\frac{2+\mu}{2}\biggl(\frac{\rho}{
\delta}\biggr)^\frac{\nu(2+\mu)}{\mu}\biggr]^{-\frac{\mu}{2+\mu}} \label{hh}
\end{equation}

where $g$ is the target gravitational acceleration, $v_\perp$ is the
perpendicular component of the impactor velocity, $\delta$ is the
projectile density, $\rho$ and $Y$ are the density and tensile
strength of the target, $k$ and $\mu$ depend on the cohesion of the
target material and $\nu$ on its porosity\footnote{The values used
  are: $k=1.03$ and $\mu=0.41$ for cohesive soils, while $k=0.93$ and
  $\mu=0.55$ for rocks.  $\nu$ has been set to 0.4 in all cases
  \citep{hol07}.}.  Therefore, the nature of the terrain affects the
crater efficiency and the functional dependence of the crater size
with respect to the input parameters (e.g. impactor size and velocity).
Equation \ref{hh} accounts both for the strength and gravity regimes,
allowing a smooth transition between the two regimes.  The impactor
size ($d_{sg}$) for which we have the transition between the two regimes
is determined by equating the two additive terms in equ.  \ref{hh},
therefore:

\begin{equation}
d_{sg}=2\frac{v_{\perp}^{2}}{g}\biggl(\frac{Y}{\rho v_{\perp}^2}\biggr)^\frac{2+\mu}{2}
\biggl(\frac{\rho}{\delta}\biggr)^\nu
\end{equation}

The transient crater diameter is converted into final crater
diameter ($D$) according to the following expressions:

\begin{eqnarray}
D & = & 1.3D_t  \hspace*{2cm} {\rm if}  \;\; D_t\leq D_{\star}/1.3 \\
D & = & 1.4\frac{D_t^{1.18}}{D_{\star}^{0.18}}  \hspace*{1.6cm} {\rm if} \;\; D_t> D_{\star}/1.3
\end{eqnarray}

where $D_{\star}$ is the observed simple-to-complex transition crater
diameter, which for Mercury is 11~km \cite{pik88}. The conversion
between transient crater to final crater is rather uncertain and
several estimates are available \citep[e.g.][]{hol93,mck85,cro85}.
Here we have used the factor 1.3 from transient to final simple
craters\footnote{The factor has been retrieved at
  http://keith.aa.washington.edu/craterdata/scaling/theory.pdf on
  January 2011.}. For the complex craters, we use the expression
proposed by \cite{cro85}, where the constant factor has been set to
1.40 in order to have continuity with the simple crater regime
($1.4=1.3^{1.18}$). We note that the effects of the material
parameters on the transient crater size depend on whether a crater is
formed in the strength or gravity regime. For the strength regime, $D
\propto \rho^{ -0.1}$ or $\propto \rho^{-0.2}$, while $D
\propto Y^{-0.2}$ or $\propto Y^{-0.3}$ according to the values for
$\mu$ and $\nu$ given above. In the gravity regime, $D \propto
\rho^{-0.3}$. In all cases, the dependence of $D$ on both $\rho$ and
$Y$ is mitigated by the low exponents.

\subsection{Inhomogeneities of the target material parameters}

\begin{figure*}[h]
\includegraphics[width=9cm,angle=-90]{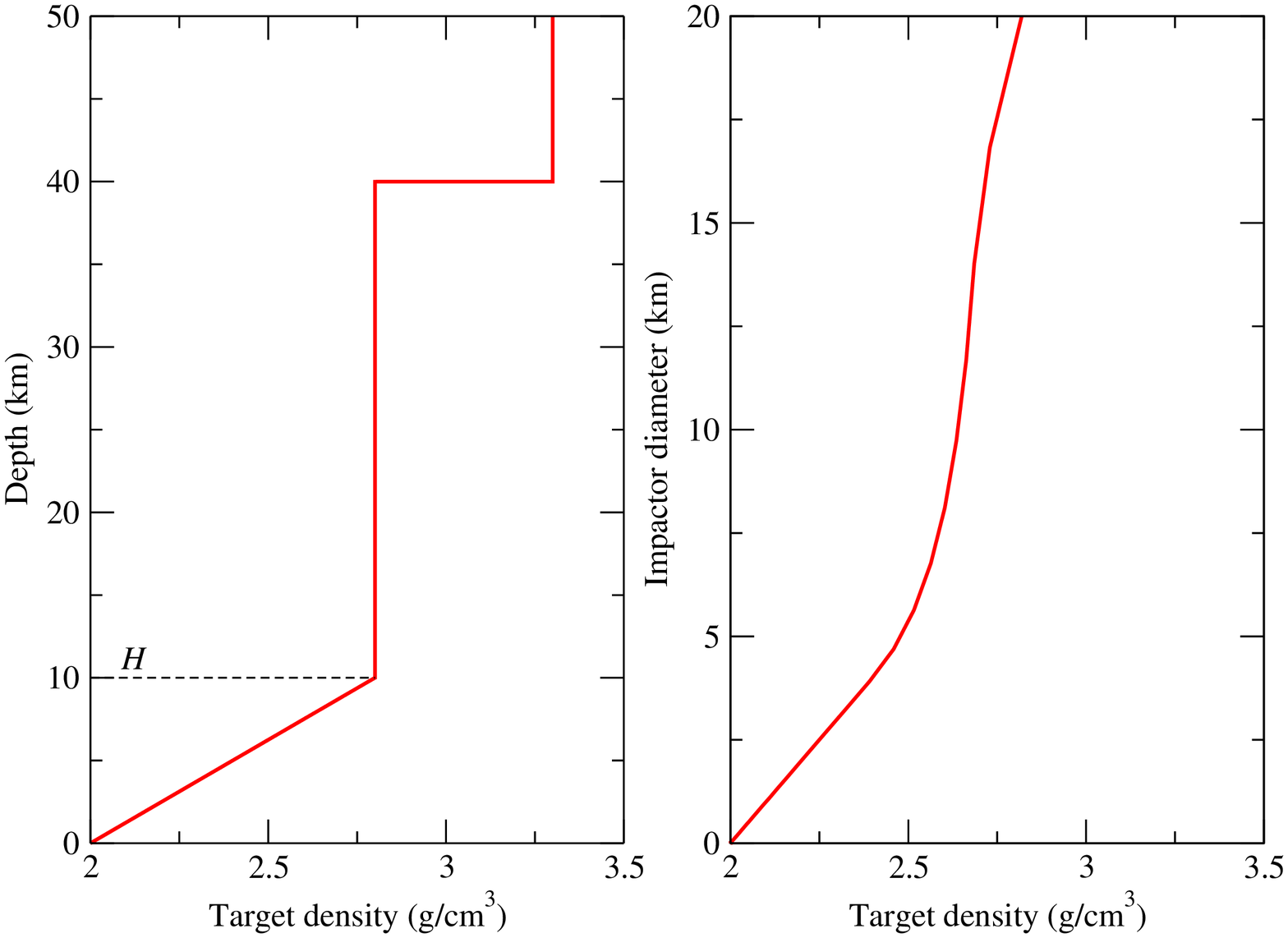}
\includegraphics[width=9cm,angle=-90]{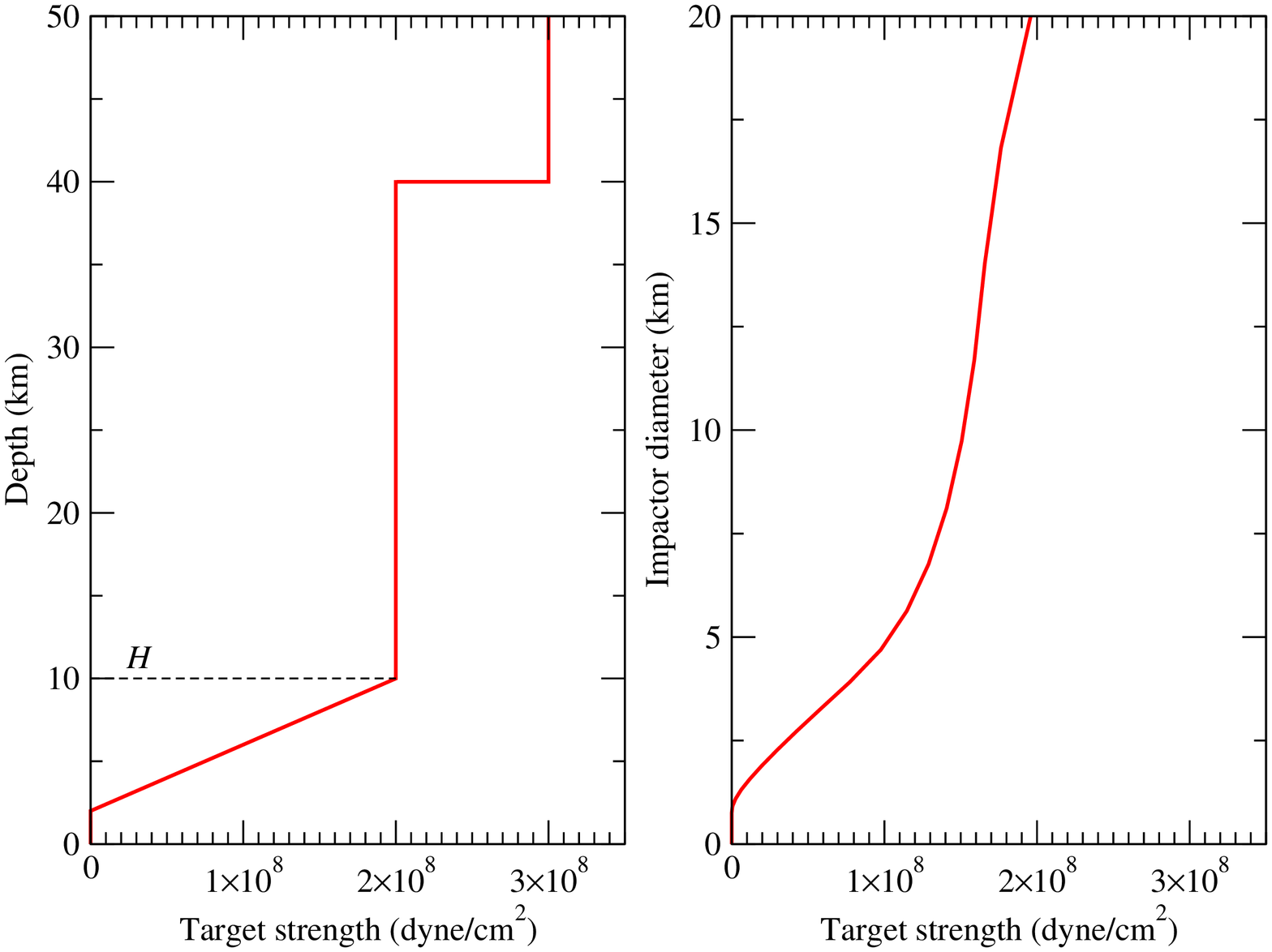}
\caption{Inferred density and strength profiles for Mercury. Left panels report
  the assumed profiles as a function of depth. The discontinuous
  transitions correspond to the major transitions in the crust, in
  analogy to the lunar crust. The depth of the upper
  cohesive layer ($H$) is also indicated for clarity. In this example,
  it has been set to 10~km. The right panel shows the averaged density
  and stregth as a function of the impactor diameter. This result has
  been achieved by averaging, for an impactor of radius $r$, both the
  density and strength up to a depth of $5r$ (see text).}
\label{den_str}
\end{figure*}

The application of the crater scaling law is not straightforward,
since the physical parameters of the terrains are poorly constrained
for Mercury and the crater scaling law has been derived for idealized
uniform target properties. So far, no detailed and systematic study
has been performed to develop a crater scaling law for a layered
target \citep[e.g.][]{mel89,col05}, although numerical modeling of
terrestrial craters has shown that the target layering plays an
important role in the cratering process \citep[e.g.][and references
  therein]{col08}.  On the other hand, we think that it is worth
attempting to simulate a more realistic situation instead of using the
same average values for craters whose sizes can vary by order of
magnitudes and consequently involved different layers of a planetary
crust.  In this context, geological analysis of the terrains can
provide valuable information, at least to constrain the surface
properties of the target.\\
In this work, we assumed that the density and strength of Mercury
varies as a function of the depth, in analogy to that inferred for the
Moon \citep[][and references therein]{mar09}. In figure \ref{den_str}
(left panels) the assumed density and strength profiles are
indicated. These profiles are consistent with the upper lunar
structure \citep{hor91}, and were adopted also for Mercury.  In
particular, we have considered a more or less fractured upper crust on
top of a bulk silicic lower crust which in turn overlays a peridotitic
mantle. However, it must be emphasized that the depths at which these
layers occur may vary from place to place (see $\S$~3).\\
For each impactor size, we have assigned average values for the target
density and strength. Over a wide range of parameters, the transient
crater radius ($R_t$) is about $10-20$ times larger that the impactor
radius $r$ and the depth of the crater is typically between one-fourth
and one-third of the crater size \citep{mel89}. Thus the thickness of
the excavated material is roughly between $2.5-7$ impactor radii.
Here we have adopted an intermediate value, namely averaging the density and
strength up to a depth of $5r$ (see fig. \ref{den_str}, right panels)
Given the limited variation in the density and strength profile, the
choice of the actual depth to average the density and strength for
a given impactor radius has a low influence ($<5\%$) on the scaling
law.\\
In addition to the density and strength profiles, we also consider a
transition of the crater scaling law (from cohesive-soil to hard-rock)
according to the size of the impactor. In fact, the density and
strength profiles shown in fig. \ref{den_str} describe a material of
increasing coherence for increasing depth. This is the result of the
continuous bombardment of planetary surfaces that produces comminution
and fracturing of the upper crustal layers gradually deacreasing with
depth as observed in seismic profiles of the lunar crust underneath
the Mare Cognitum \citep{tok72,sim73,kha00}. In this respect, craters
that affect only the upper fractured layers form in the cohesive-soil
regime, while larger ones in hard-rock regime. Therefore the depth of
the transition ($H$) from the superficial fractured layer to the
unfractured lower crust is an important parameter.  The depth (and
therefore the crater size) at which the transition from one regime to
the other occurs can vary from place to place. For instance, the
thickness of the cohesive layer may be only of a few meters on recent
lunar mare material \citep{qua68}, while it is expected to be of
several kilometers on the highlands\footnote{Note that, in principle,
  the above scenario maybe locally reversed, e.g. in the presence of
  fresh solidified lava on top of an older fractured layer.}.  In the
examples of fig.  \ref{den_str}, it is assumed that $H=10$~km
\citep{hor91}.\\
The details of the transition are not easy to model.  A simplified
study of impact processes on a layered target was performed by
\citep{qua68}.  They simulated a two-layer structure, formed by a
loose, granular layer on top a more competent material. It was
observed that the craters had the usual shapes for diameters less than
about 4 times of the top layer thickness. Larger craters developed
central mounds, flat floors and concentric rims indicating the
presence of the underlying layer.  According to these results, we
simplify the problem by considering a sharp transition in the crater
scaling law, at $D_t=4H$. This implies a transition as a function of
impactor radius at $r=H/5-H/10$.  The effect of the transition in the
scaling law is reported in fig. \ref{SL} for two depths of
transitions. Note that the position of the sharp transition varies
according to the depth assumed.  In this paper, we use an intermediate
value and set the transition at $r=H/7.5$.  In a more realistic
situation, a gradual transition should be predicted given that the
target gradually changes its properties as function of the depth.
Therefore, our simplified model is not expected to be accurate close
to the transition region, nevertheless we believe it provides a
reasonable way to approach the cratering scaling law for a layered
target.\\

\subsection{Deriving the  Model Production Function}

\begin{figure*}[h]
\includegraphics[width=12cm,angle=-90]{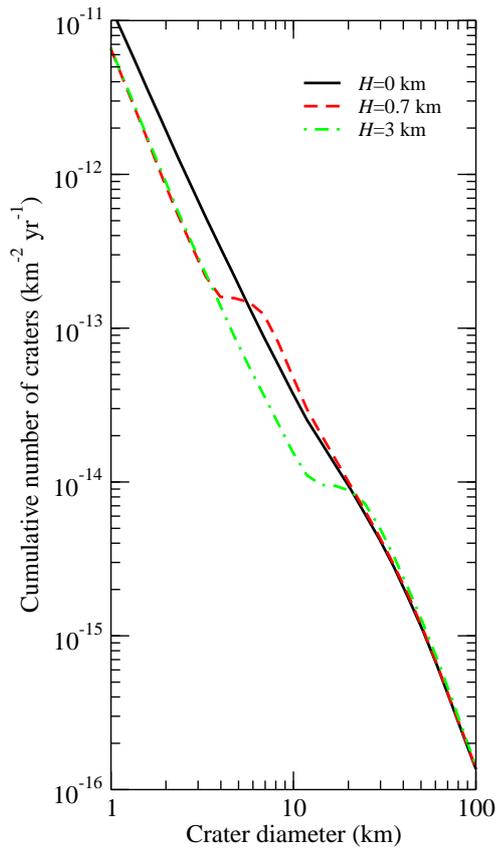}
\caption{Examples of computed crater size-frequency diatributions per
  unit surface per unit time (the so-called Model Production Function,
  MPF) obtained with three different values of $H$, which will be used
  for the age assesment of Rachmaninoff and Raditladi basin.  In all
  cases, the transition in the scaling law is set at $r=H/7.5$.
  Depending on the relative position of the S-shaped feature with
  respect to the crater SFD, a maximum age variation of a factor 3-4
  can occur.}
\label{mpf}
\end{figure*}

The NEO population and the crater scaling described in the previous
section are used to derive the MPF per unit time (see fig.
\ref{mpf}).  The main outcome of our model is that the adopted
transition in the crater scaling law results into a ``S-shaped''
feature (or flexure) in the MPF.  The position of such feature, which
is determined by $H$, is not known a-priori.  However, as discussed in
\cite{mas09}, in some cases $H$ can be constrained by the shape of the
observed crater SFD.  Furthermore, the geological analysis of the
terrains can help to derive the expected range of variation for $H$.
For instance, lava emplacements may partly strengthen or even
completely replace the pre-existing fractured layer. Hence in this
latter cases the fractured horizon can be confined within a very thin
regolith cover, negligible for our calculation ($H\sim0$).  For young
units with poor crater statistics, the choice of $H$ may affect the
age determinations by up to a factor of 3-4. Thus, in order to derive
a more accurate age estimate, it is of paramount importance to adapt
the crater production function to the nature of the terrains
investigated.\\
The absolute age is given by the
lunar chronology, which expresses the lunar crater cumulative
number at 1~km ($N_1$) as a function of time ($t$), using the
following equation:

\begin{equation}
N_1(t)=a(e^{bt}-1)+ct \label{chrono_eq}
\end{equation}

where $a=1.23\times10^{-15}$, $b=7.85$, $c=1.30\times10^{-3}$
\citep{mar09}. The MPF function at a time $t$ is given by:

\begin{equation}
{\rm MPF}(t)={\rm MPF}(1{\rm yr})\cdot\frac{N_1(t)}{N_1(1{\rm yr})} \label{mpf_eq}
\end{equation}

The MPF($t$) is used to derive the model cratering age by a best fit
procedure that minimizes the reduced chi squared value, $\chi_r^2$.
Data points are weighted according to their measurement errors.  The
formal errors on the best age correspond to a 50\% increase of the
$\chi_r^2$ around the minimum value.\\
It must be realized that the formal statistical error on the model age
only reflects the quality of the crater SFDs. On top of that, other
sources of uncertainties are present.  They stem from the
uncertainties involved in the physical parameters used in the model,
although we want to stress here that the model ages are not very
  sensitive to details of the density and strength profile \citep[see
    $\S$~3.1 of][]{mar09}.  A more important issue is the
  applicability of the present NEO population in the past. The
  $N_1(t)$ chronology function assumes a linear dependence with time
  in the last $\sim 3.8$~Ga, corresponding to an impactor population
  in a steady state.  On the other hand, dynamical studies of recent
  main belt asteroid family formation suggested that the present NEO
  flux may be higher that the average steady flux by a
  factor of 2 \citep{bot07}. This result also agrees to what
  found by cratering studies of young lunar terrains
  \citep[$<0.8$~Ga; e.g.][]{mar09}.
Concerning the layering structure, as described above, the choice of
$H$ might affect the age estimate by a factor of 3-4 at most. The
uncertainty due to the layering is, nevertheless, typically present
only for young terrains where the crater SFD has a limited range of
crater dimensions.  The layering affects the specific shape of the
crater SFD, and has to be evaluated case by case, as it will be
discussed in detail in the following sections. Finally, it must
  be noted that wavy features in the crater SFD can be due also to
  other processes than layering, like for instance partial crater
  obliteration due to subsequent lava flows. Therefore, the nature of
  the S-shaped feature must be constrained as much as possible by
  geological analysis in order to achieve a more reliable age
  determination.

\section{Geological analysis and model ages}

Geological maps of the Rachmaninoff and Raditladi basins were
constructed considering both floors and ejecta.  For the floor
terrains, the geological units were identified on the base of their
different surface morphologies and spectral characteristics (i.e.,
albedo), along with an analysis of their stratigraphic relationships.
The ejecta units, surrounding the basins, were outlined considering
exclusively the area of continuous ejecta blankets, which are easily
detectable thanks to their characteristic hummocky surface.  The
geological maps also take into account tectonic features affecting the
areas.\\
Crater age determination is based on the primary craters, i.e. those
formed by impacts with objects in heliocentric orbits.  Hence, a
crucial point in assessing age by crater counts is to identify and
avoid secondary craters. Most of the secondaries are recognizable
because they are directly related to their primaries (e.g., the
secondaries are arranged in radial patterns around the primary), or
occurr in loops, cluster and chains.  The contribution of far-field
secondaries, which are normally not distinguishable from primary
craters \cite[e.g.][]{mce06}, has been neglected. Although this is a
reasonable assumption for Rachmaninoff and Raditladi basins given
their low crater densities and thus their presumable relatively young
ages, MPF model ages may overestimate real ages.\\

\subsection{Rachmaninoff basin}

\begin{figure*}[h]
\includegraphics[width=10cm]{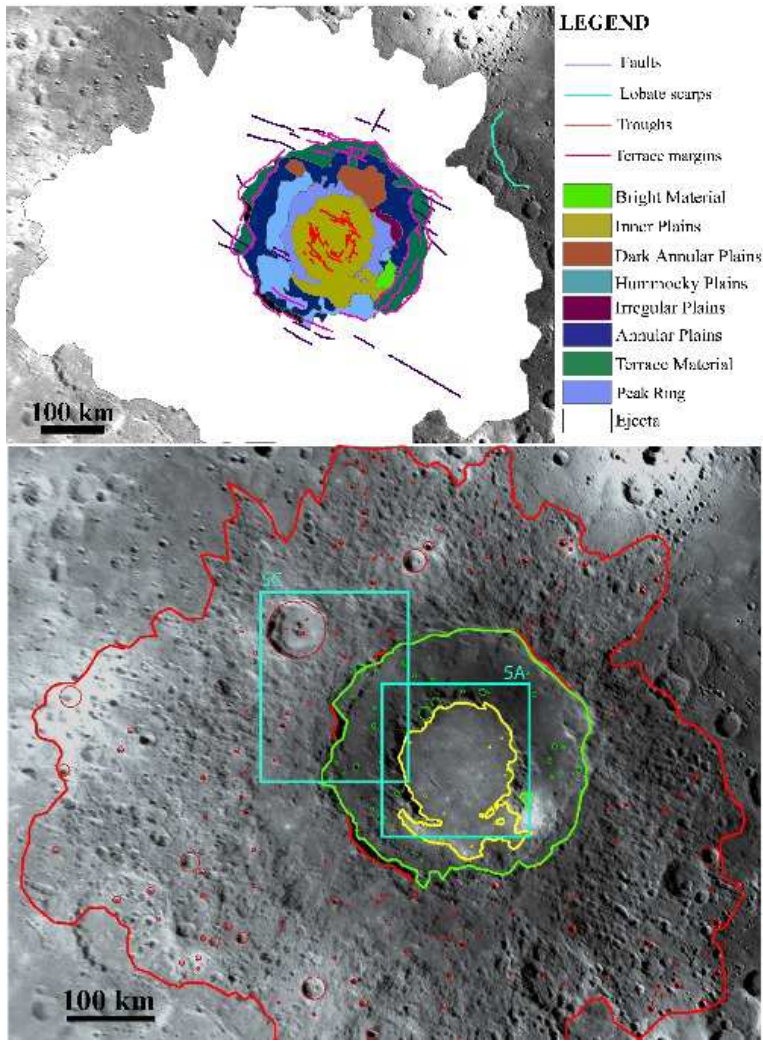}
\caption{Upper panel: Rachmaninoff geological sketch. Lower panel:
  Bonafide craters within the three regions considered for crater
  counts. Boundaries and bonafide craters are in yellow for the inner
  plains, green for the annular units and red for the ejecta
  blankets. The counting areas are: 1.74$\times 10^4$, 4.67$\times
  10^4$ and 3.32$\times 10^5$ km$^2$ for inner plains, annular units
  and ejecta blanket, respectively. Boxes indicate close views of
  fig. \ref{rach_ex}.}
\label{rach_geo}
\end{figure*}

\begin{figure*}[h]
\includegraphics[width=10cm]{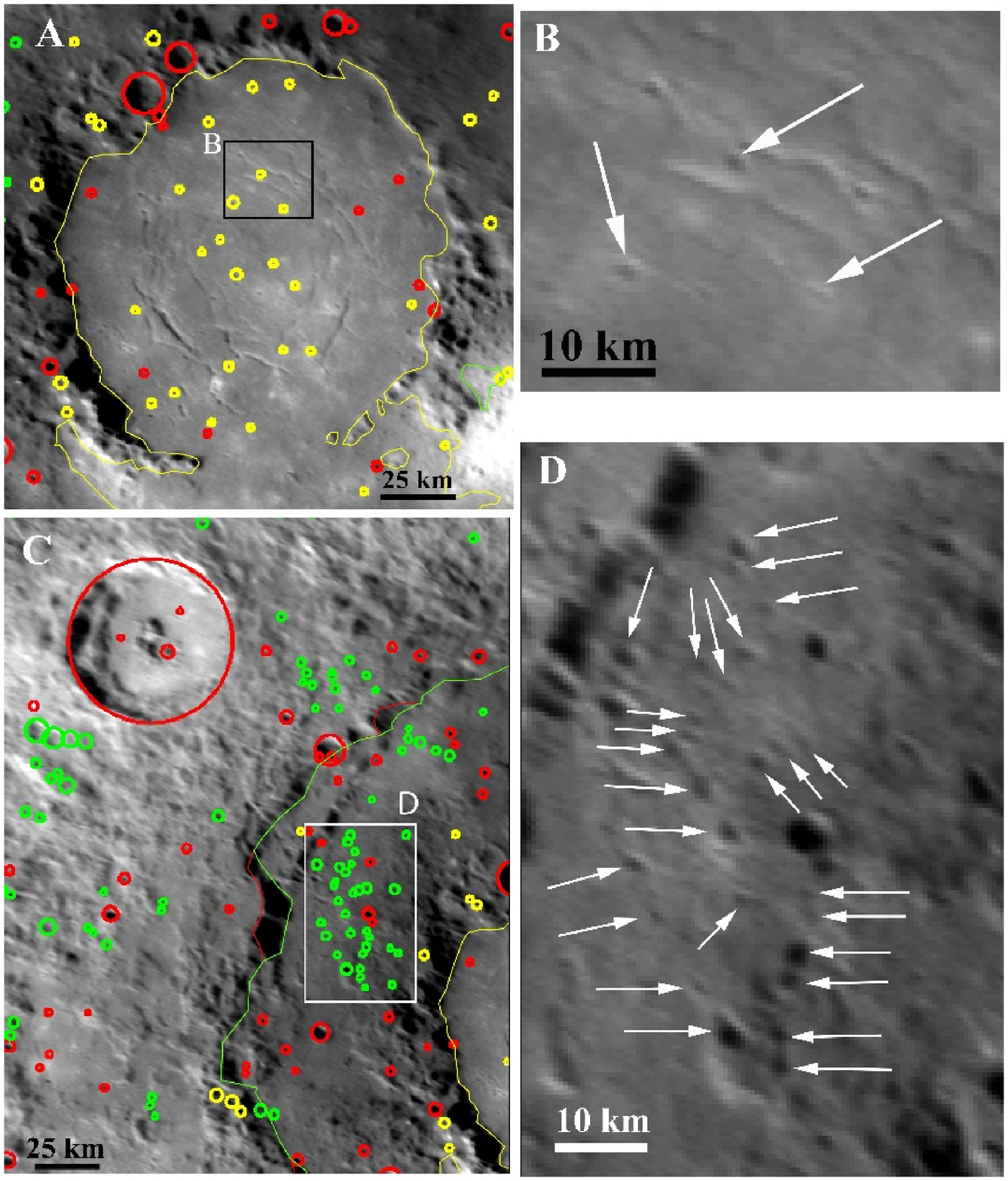}
\caption{A) Close view of the plains within the Rachmaninoff peak
  ring: red circles = bonafide craters, yellow circles = circular
  features of endogenic origin; green circles = secondary craters. B)
  Detail of endogenic features (white arrows), often with irregular
  shapes and generally related to troughs and/or surrounded by
  brighter aloes. C) Close view of a primary peak crater (60~km of
  diameter) and related secondary craters: red cicles = bonafide
  craters, yellow circles = circular features of endogenic origin;
  green circles = secondary craters . D) Detail of secondary craters
  (white arrows) often with elliptical shapes and associated in chains
  and loops.}
\label{rach_ex}
\end{figure*}

\begin{figure*}[h]
\includegraphics[angle=-90,width=9cm]{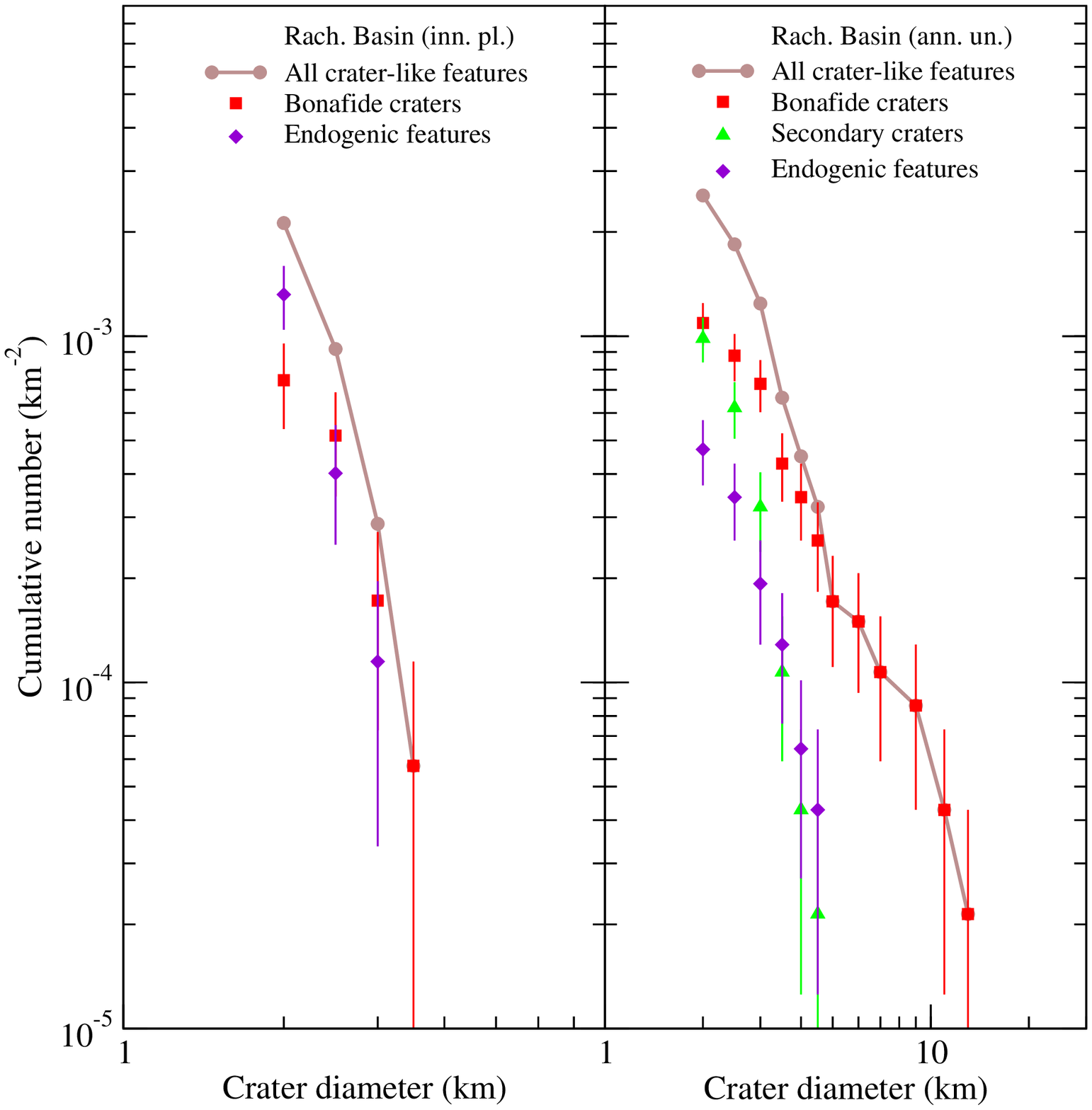}
\includegraphics[angle=-90,width=9cm]{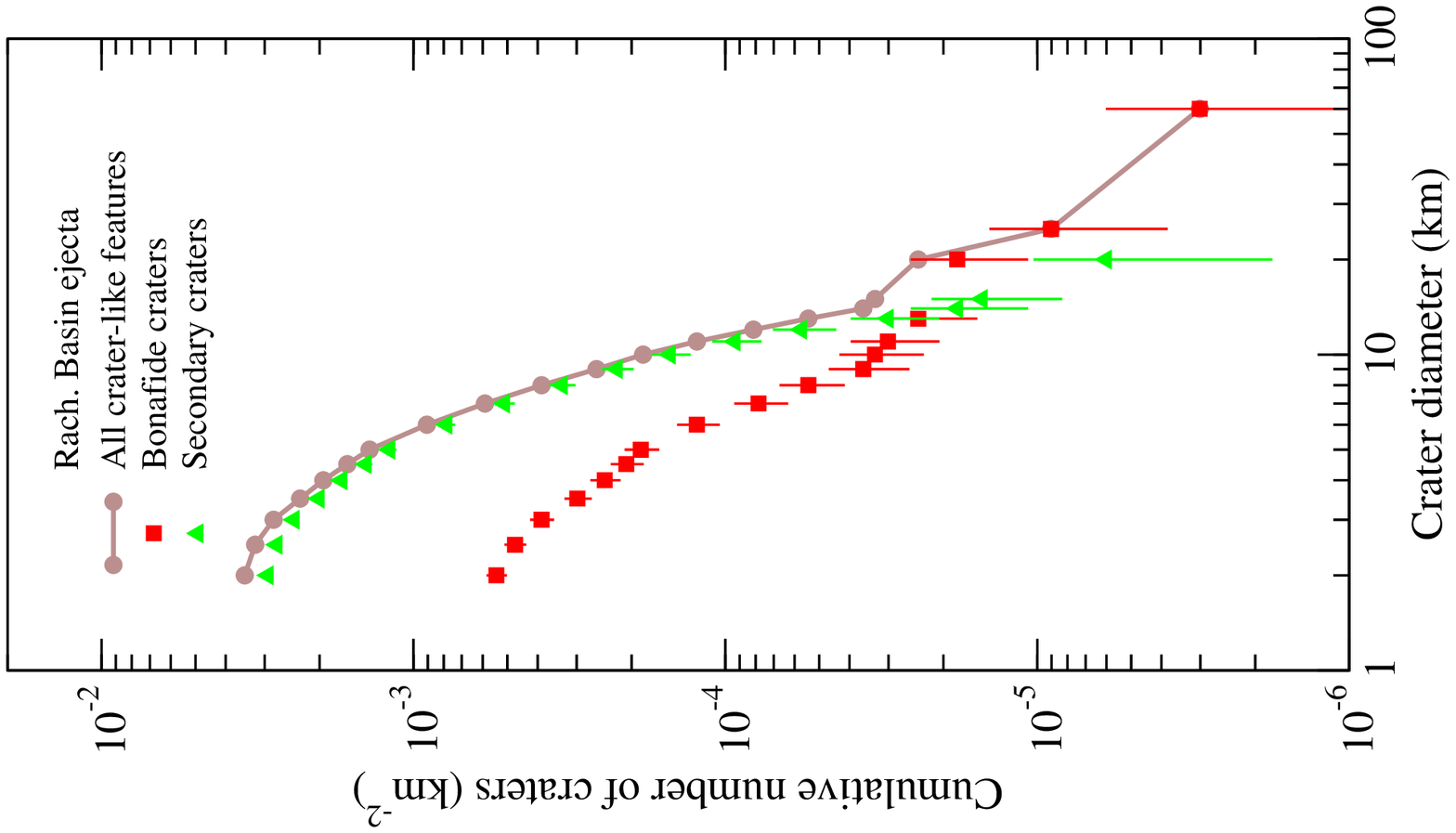}
\caption{Rachamaninoff basin SFDs of the detected features for inner
  plains and annular units (upper panels) and ejecta (lower panel).}
\label{rach_SFD}
\end{figure*}

\begin{figure*}[h]
\includegraphics[angle=-90,width=9cm]{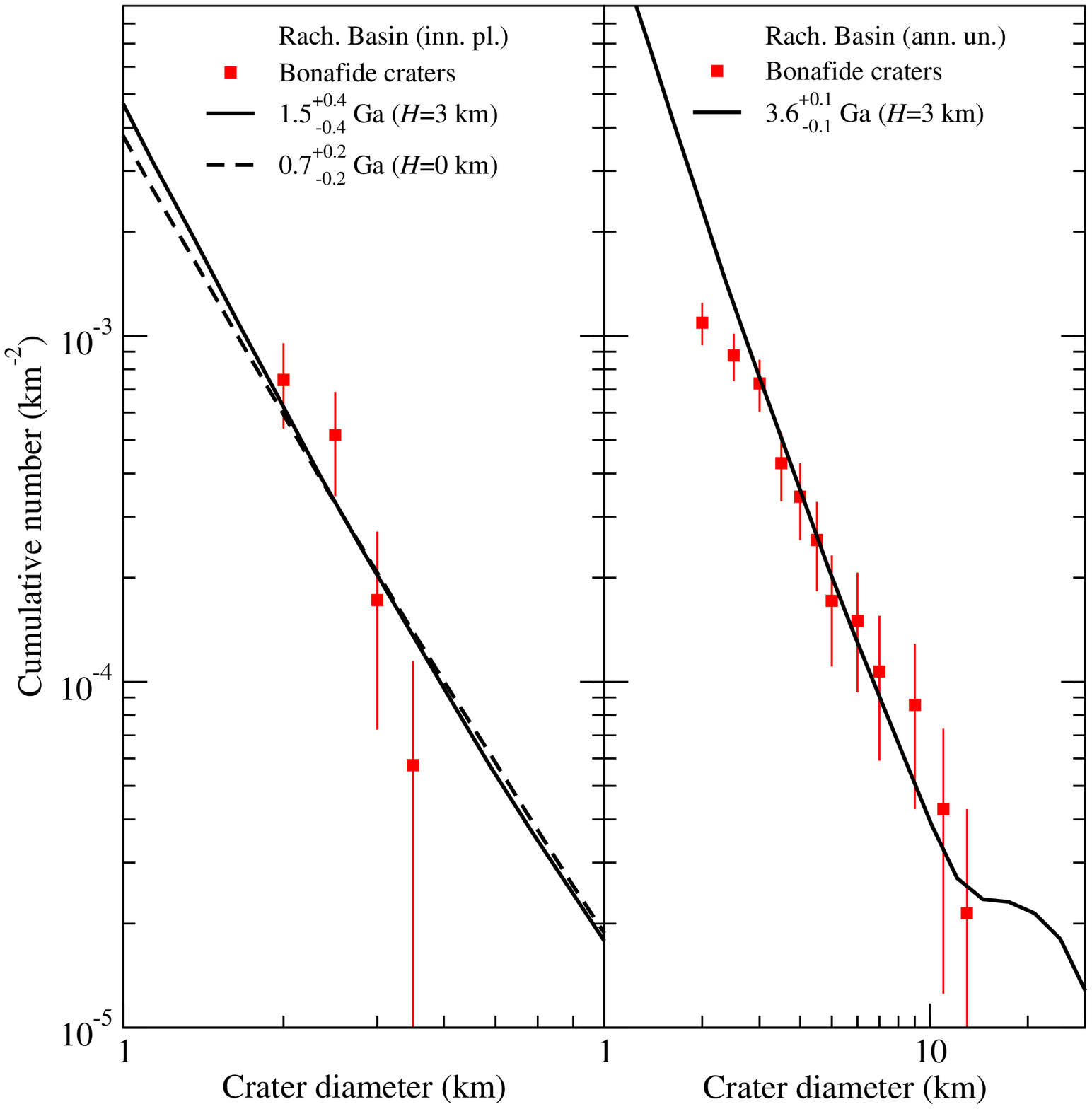}
\includegraphics[angle=-90,width=9cm]{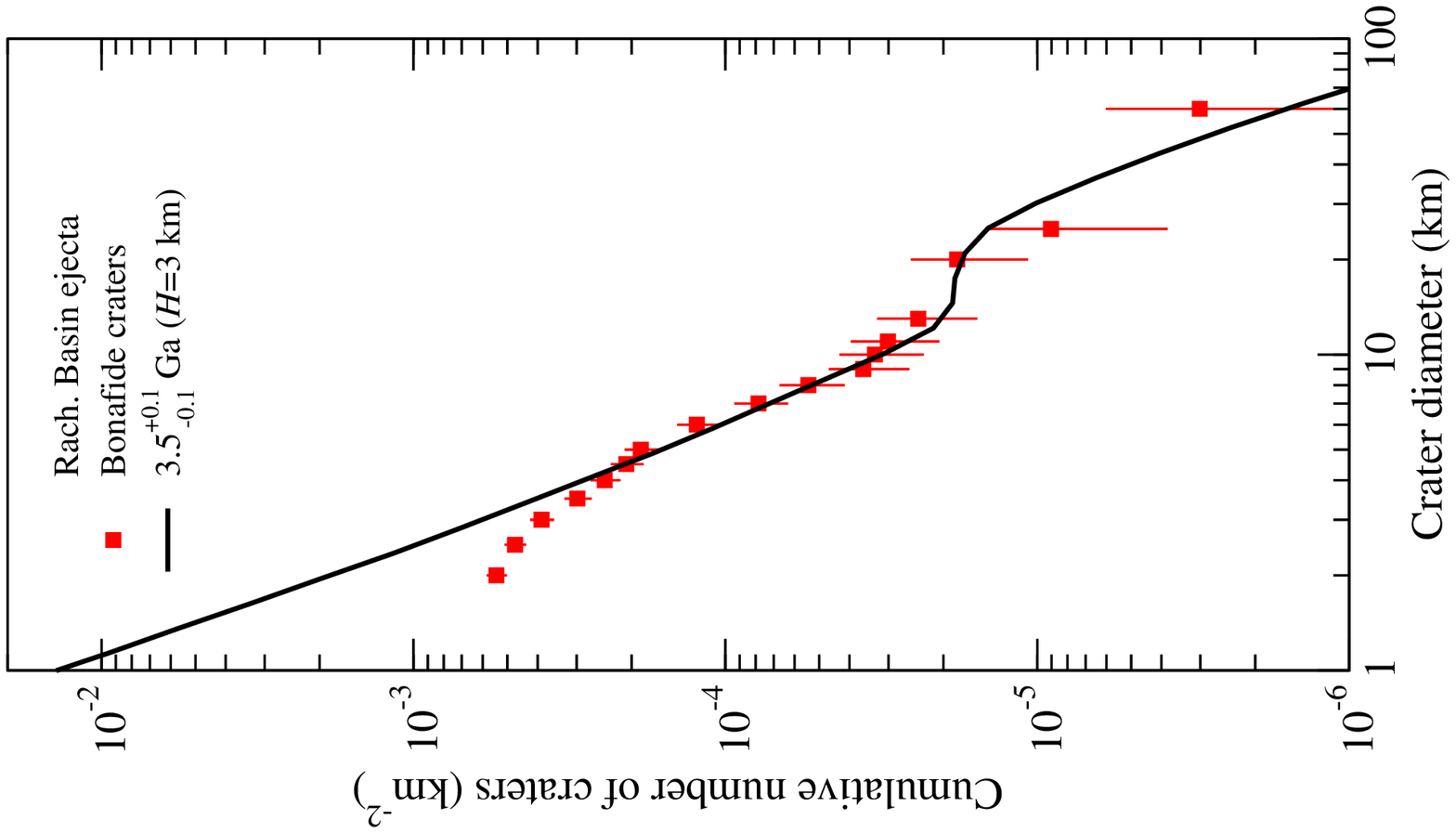}
\caption{Rachmaninoff age assessment. MPF best fit of the bonafide
  crater SFDs for the inner plains and annular units (upper panels)
  and ejecta (lower panel).}
\label{rach_MPF}
\end{figure*}

\begin{figure*}[h]
\includegraphics[width=12cm]{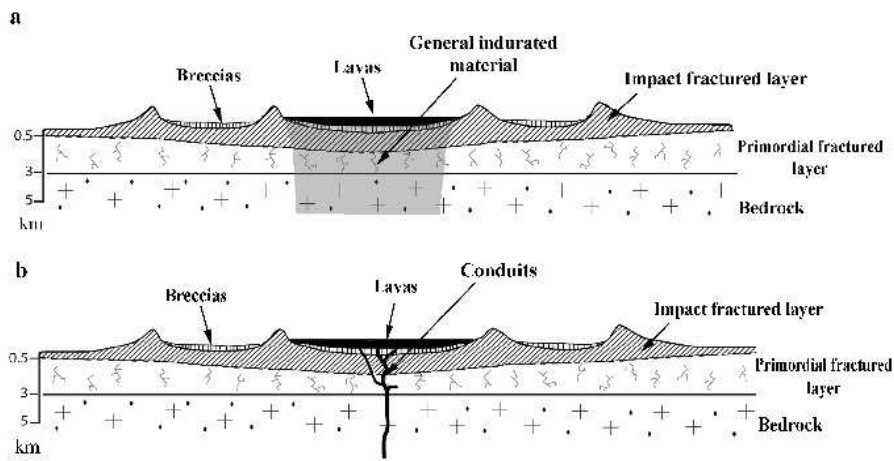}
\caption{Possible geological sections of the Rachmaninoff basin
  hypothesized from the fit of MPF with the crater SFD of ejecta,
  annular materials and inner plains: a) rising magmas and lavas
  completely hardened the former horizon made up of primordial and
  impact related fractured material ($H=0$~km); b) a weakly sustained
  volcanism emplaced a thin lava layer on top of the fractured
  material with magma influx concentrated along few conduits
  ($H=$0.7~km).}
\label{rac_scenario}
\end{figure*}

The Rachmaninoff basin is surrounded by a continuous ejecta blanket and
includes an interior peak ring structure, about 136~km in diameter,
with extended smooth plains filling its floor. Most of the basin walls
are modified into terraces.  Several different geological units have
been distinguished inside the floor on the basis of their different
relative albedo and surface texture (fig.  \ref{rach_geo}). 
  The inner smooth plains are mostly within the peak ring except in
  the southern quadrant of Rachmaninoff, where the smooth plains cover
  or embay the peak ring structure and some of the annular units within the
  rim.  This observation suggests an origin of volcanic emplacement. In
the WAC enhanced-color images these plains show a yellow to reddish
tone which stands out from the darker and bluer color of the other
units within the basin and surrounding regions \citep{pro10}. This
clearly supports a different composition and origin of these inner
smooth plains.  Several discontinuous and concentric troughs possibly
due to the uplift and extension of the basin floor, affect the area
enclosed by the peak ring \citep{pro10} and have been interpreted to
be graben.  The annular region between the peak ring and the rim basin
includes seven different units. The most prominent is made up of
bright materials, apparently younger than all the other units and
possibly related to explosive volcanism \citep{pro10}.  Peak ring and
terrace material boundaries stand out for their relief, whereas
hummocky, dark, irregular and annular smooth plains do not show
unequivocal stratigraphic relationships with each other \citep{pro10}.
This suggests an almost coeval origin of these units that may
consist of impact melts and breccias. This is furthermore confirmed by
the WAC images, where annular units do not reveal any color variations
and are characterized by  uniform blue color similar to the
surrounding terrain.  To shed more light on the origin of the floor
material we dated the Rachmaninoff basin using crater statistics of
annular units and inner plains separately.  Bright material was
neglected in the crater counts due to its limited extent. Craters in the
ejecta blanket were counted as well (fig. \ref{rach_geo}).\\
Easily recognizable secondary craters (either elliptical in shape
  or arranged in loops and chains) have not been detected within the
Rachmaninoff inner plains but numerous pits up to 3.5~km in diameter
have been found in close proximity to the concentric grabens.  These
features are very unlikely to be impact craters and in our
interpretation are most probably of tectonic  (structural pits,
  fault bounded depressions, en-echelon structures) and/or volcanic
  (more or less irregular vents) origin (fig. \ref{rach_ex}, panels A
  and B).  For this reason their counts were neglected for the
  purposes of age determination.  Clusters and chains of secondaries
with irregular and elliptical shapes were recognized in the western
sector of the annular units  and appear to be directly related to
  a nearby 60 km primary peak crater overlying the Rachmaninoff ejecta
  (fig. \ref{rach_ex}, panels C and D). Self secondaries are numerous
within the ejecta blanket.\\
The resulting crater count statistics are reported in Table
  \ref{tab} where for each terrain, all crater-like features, bonafide
  craters, secondary craters and endogenic (namely  volcanic or
  tectonic) features are listed.  It is interesting to compare the
  SFDs of all the counts (fig. \ref{rach_SFD}).  According to our
  best interpretation of the detected features, the inner plains
  contain more endogenic features than bonafide craters.  Hence, in our
  opinion the uneven distribution in R plots of crater-like features
  smaller than 4~km, which is generally attributed to the effect of
  far field secondaries on Mercury \citep{str08}, are in this case
  dominated by tectonic and volcanic features.  For the annular units,
  both the secondary crater and endogenic SFDs have steeper slopes with
  respect to the bonafide crater SFD, moreover they are limited to
  features smaller than 4-5~km.  Hence, for the annular units as well
  as the inner plains, the identification of endogenic features is
  clearly very important since it heavily affects the final bonafide
  crater SFD.  On the ejecta, most of the crater-like features appear
  to be self secondaries mostly arranged into clusters and chains
  and/or with an elliptical shape. All the terrains were dated using
  their bonafide crater SFDs.\\
\begin{table}
\caption{ Statistics of all the features detected on Rachmaninoff and
  Raditladi basin floors and ejecta.  ``All'' indicates all
  crater-like features, ``Bon'' the bonafide craters, ``Sec''
  secondary craters (which includes chain, cluster and elliptical
  craters), ``End'' endogenic (volcanic and tectonic) features.
  ``Inn. pl.'' and ``Ann. un.'' stand for inner plain and annular
  unit, respectively.  }
\label{tab}
\begin{tabular}{l|ccc|ccc}
\hline 
\hline 
& \multicolumn{3}{c}{Rach} & \multicolumn{3}{c}{Rad} \\ \cline{2-7} 
Count & Inn. pl. & Ann. un. & Ejecta & Inn. pl. & Ann. un. & Ejecta \\ \hline
All & 37 & 119 & 1154 & 231 & 214 & 1518 \\ 
Bon & 13 & 51 & 180 & 79 & 96 & 452 \\ 
Sec & 1 & 46 & 974 & 34 & 91 & 1029 \\ 
End & 23 & 22 & 0 & 118 & 27 & 37 \\ 
\hline 
\hline
\end{tabular}
\end{table}
The MPF fits of the observed crater SFDs are shown in figure
  \ref{rach_MPF}.  The lower panel shows the distribution of bonafide
  primary impact craters detected on the ejecta blanket.  A remarkable
  feature in the crater SFD of the ejecta is the presence of a flexure
  point at about $D=15$~km. The actual shape of the bonafide crater
  SFD is partially due to the feature selection.  Nevertheless, we
  think that at the large crater sizes relevant here, our selection is
  reliable and, consequently, the observed flexure point is likely a
  real feature possibly reflecting a layered target with an upper weak
  horizon.  Hence the MPF best fit is achieved with $H=3$~km and gives
  a model age of $3.54\pm0.1$~Ga.\\
  In fig. \ref{rach_MPF} (upper panel) the bonafide crater SFD on
  inner plains and annular units are shown.  Note that, unlike the
  ejecta, both cases do not show the presence of a flexure point.
  This may be due to a real absence of an upper weak horizon or to the
  lack of large craters that would have otherwise allowed
  to retrieve information on the geomechanical properties of the deep
  crustal layers. The annular units are composed of breccias more or
  less welded by impact melts which can have only partially
  strengthened the fractured material either pre-dating or originating
  from the Rachmaninoff impact. Hence, it is reasonable to assume at
  least the same $H$ of the crust beneath the ejecta. This leads to a
  model age of $3.6\pm0.1$~Ga, which is consistent with the model age
  of the ejecta and likely dates the Rachmaninoff impact event.\\
The inner plains are characterized by much poorer statistics within a
small range of diameters, therefore the crater SFD cannot be used to
infer $H$. Nevertheless, geological analysis suggest that the inner
plains are younger volcanic flows on the basis of their different
albedo, color and overlapping relationship with respect to the unit
emplaced between the peak-ring and the basin rim \citep{pro10}.  This
would make possible also the scenario in which the former fractured
horizon, either pre-dating or originating from the impact itself, was
completely hardened by the rising magmas and emplacement of lava
fields (fig. \ref{rac_scenario}a).  In this case, the MPF acceptably 
fits the bonafide crater
size distribution giving a model age of $0.7\pm0.2$~Ga
(fig. \ref{rach_MPF}).  By contrast the fit would be very poor if the
all crater-like features would be taken into account.  This is not
surprising considering the strong contribution we infer that tectonic
and volcanic features have on the inner plains statistics.  Another
possible scenario is that the magmatic activity within Rachmaninoff
was unable to totally strengthen the upper weak layer comprising of
fractured material originated by the impact itself or inherited by
primordial events. This could be due to either weakly sustained
volcanism, which emplaced a thin volcanic sequence on top of a
fractured material, and/or, a magma influx concentrated along few well
defined conduits within a still fractured crust underneath the
basin (fig. \ref{rac_scenario}b). In this case we also computed 
the model age using $H=3$~km as
for the annular units and ejecta, obtaining a value of $1.5\pm0.4$~Ga.
In both cases (upper weak crustal layer absent or preserved) the inner
plains turn out to be remarkably young, and demonstrate that a recent
volcanic activity occurred within the basin.\\

\subsection{Raditladi basin}

\begin{figure*}[h]
\includegraphics[width=10cm]{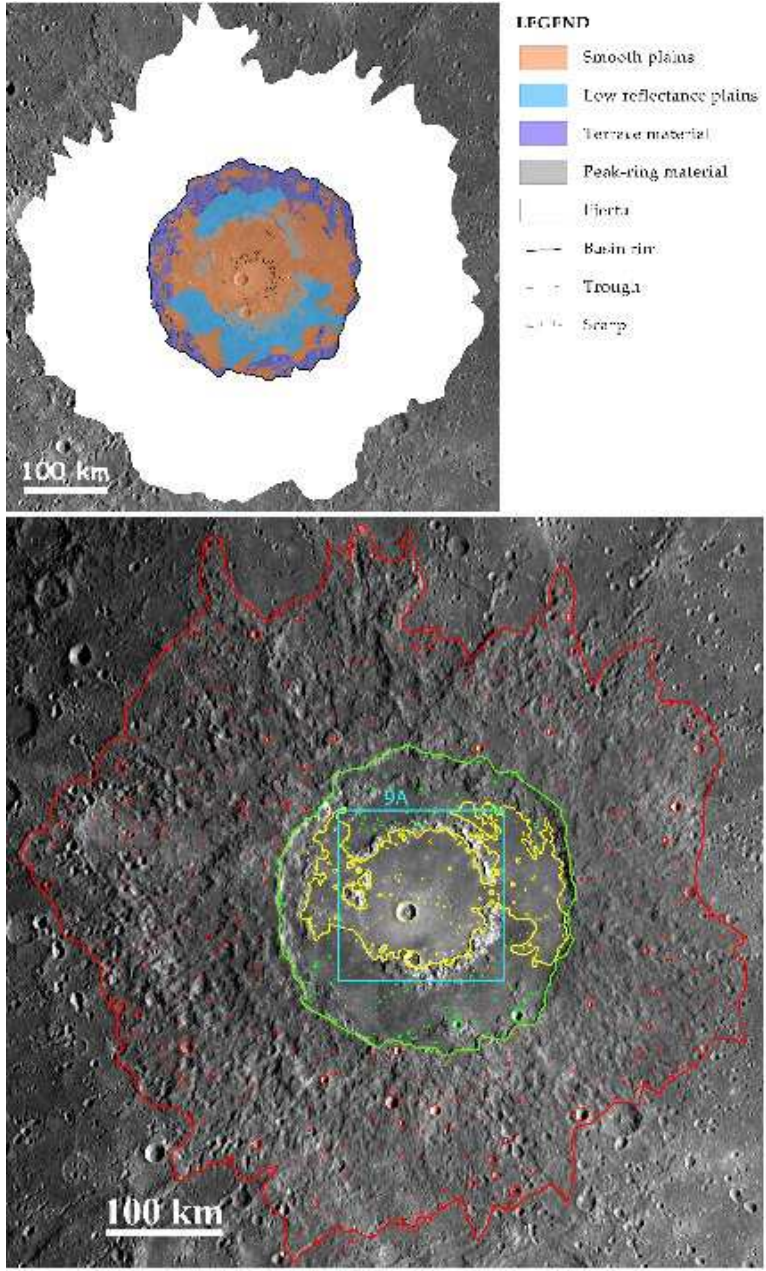}
\caption{Upper panel: Raditladi geological sketch. Lower panel:
  Bonafide craters within the three regions (inner plains, annular
  units, and ejecta blanckets) considered for crater
  counts. Boundaries and bonafide craters are in yellow for the inner
  plains, green for the annular units and red for the ejecta
  blankets. The counting areas are: 1.95$\times 10^4$, 3.39$\times
  10^4$ and 2.01$\times 10^5$ km$^2$ for inner plains, annular units
  and ejecta blanket, respectively.  Boxes indicate close view of
  fig. \ref{rad_ex}.}
\label{rad_geo}
\end{figure*}

\begin{figure*}[h]
\includegraphics[width=10cm]{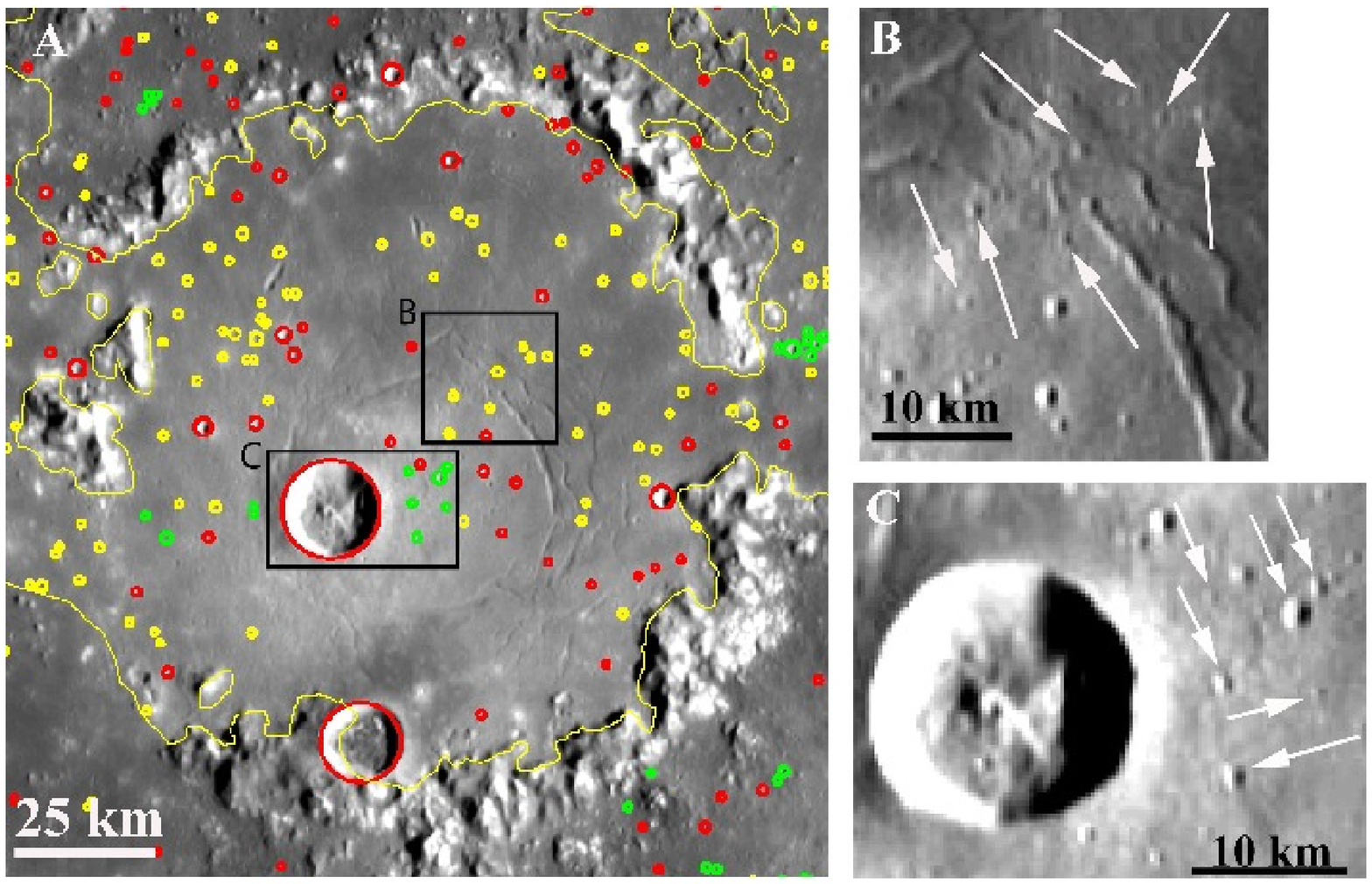}
\caption{A) Close view of the plains within the Raditladi peak ring:
  red circles = bonafide craters, yellow circles = circular features
  of endogenic origin; green circles = secondary craters. B) Detail of
  endogenic features (white arrows), often with irregular shapes and
  generally related to troughs and/or surrounded by brighter aloes. D)
  Close view of a primary peak crater (18~km of diameter) and related
  secondary craters (white arrows).}
\label{rad_ex}
\end{figure*}

\begin{figure*}[h]
\includegraphics[angle=-90,width=9cm]{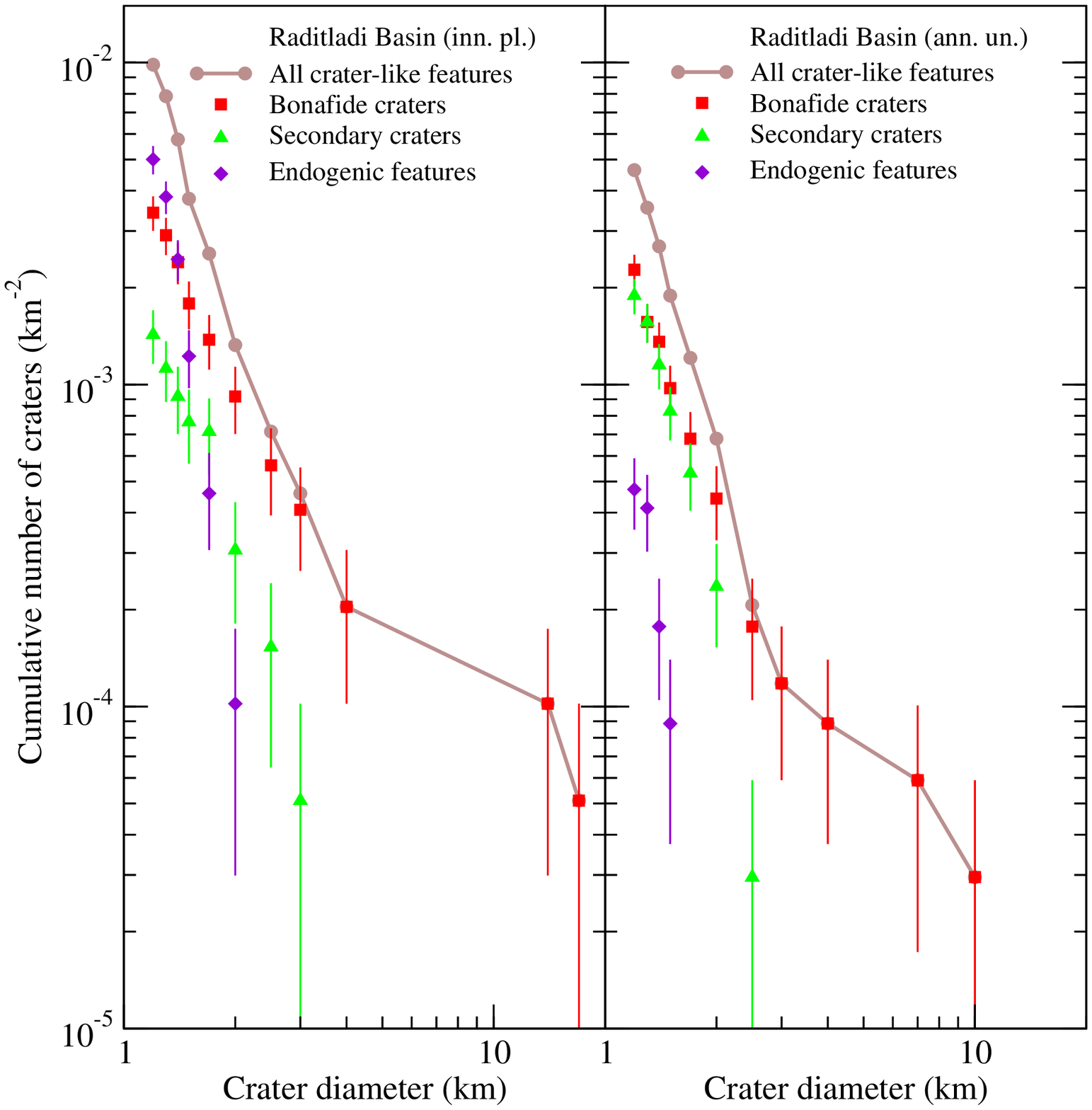}
\includegraphics[angle=-90,width=9cm]{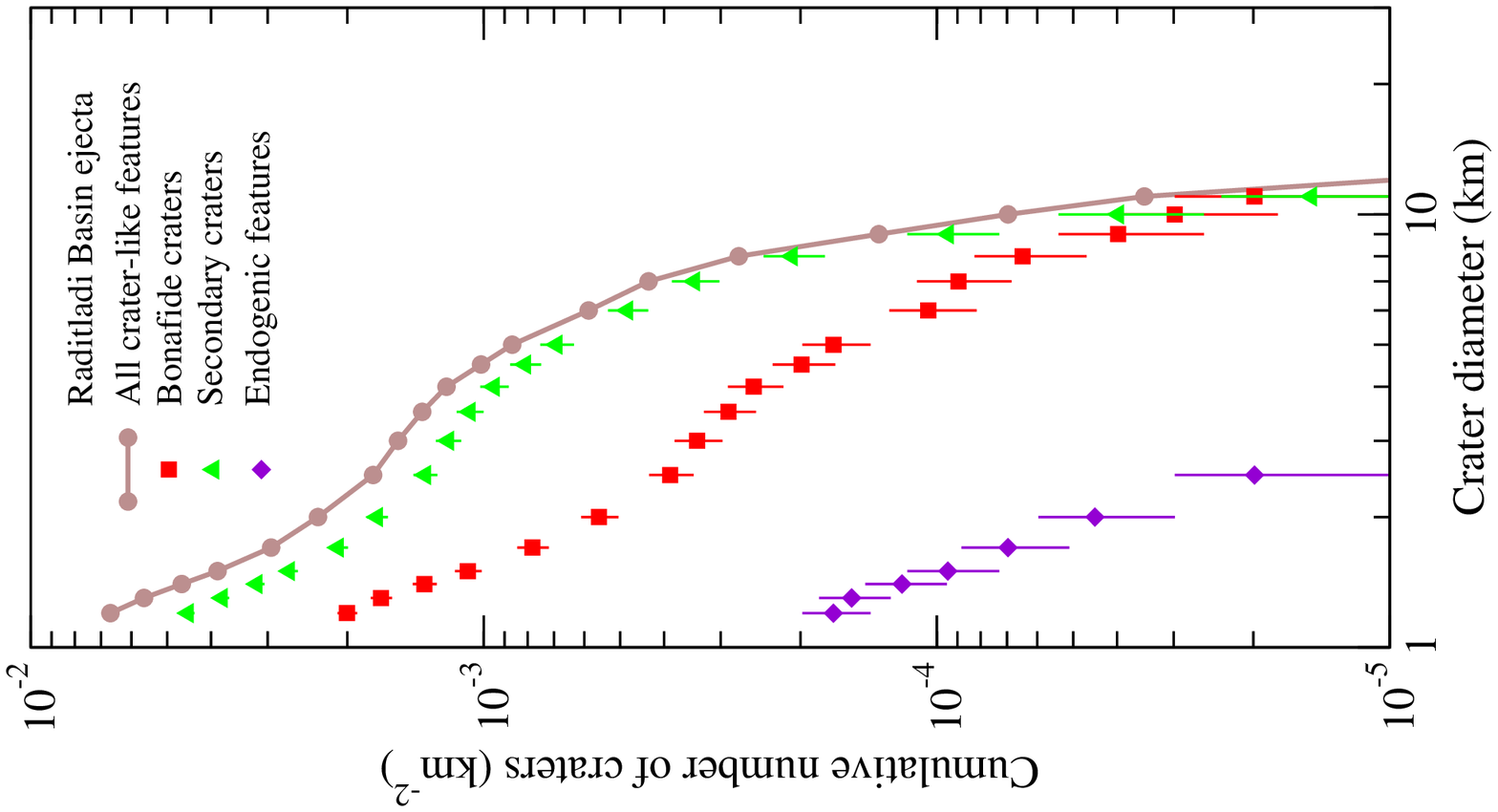}
\caption{Raditladi basin SFDs of the detected features for inner
  plains and annular units (upper panels) and ejecta (lower panel).}
\label{rad_SFD}
\end{figure*}

\begin{figure*}[h]
\includegraphics[angle=-90,width=9cm]{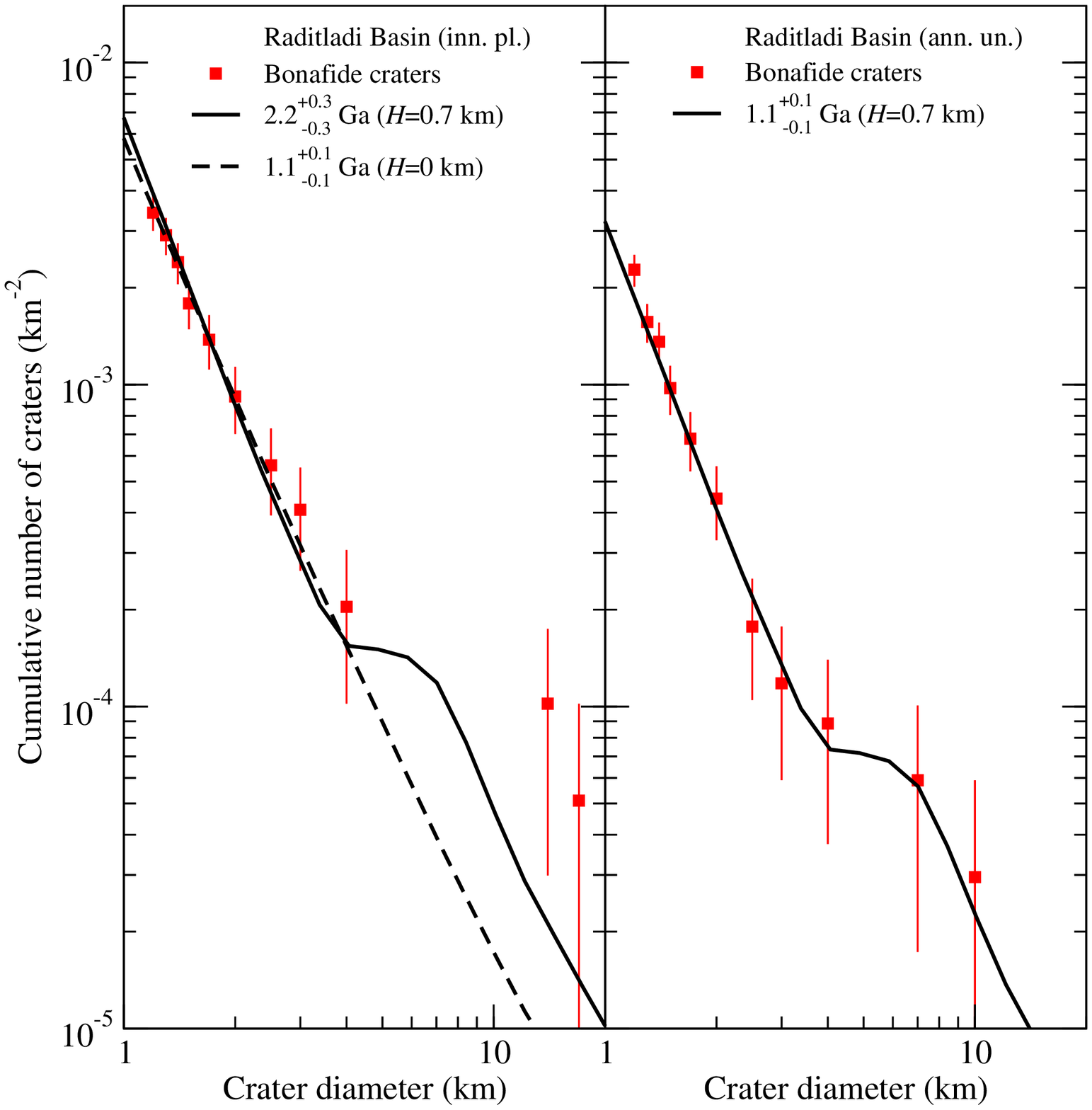}
\includegraphics[angle=-90,width=9cm]{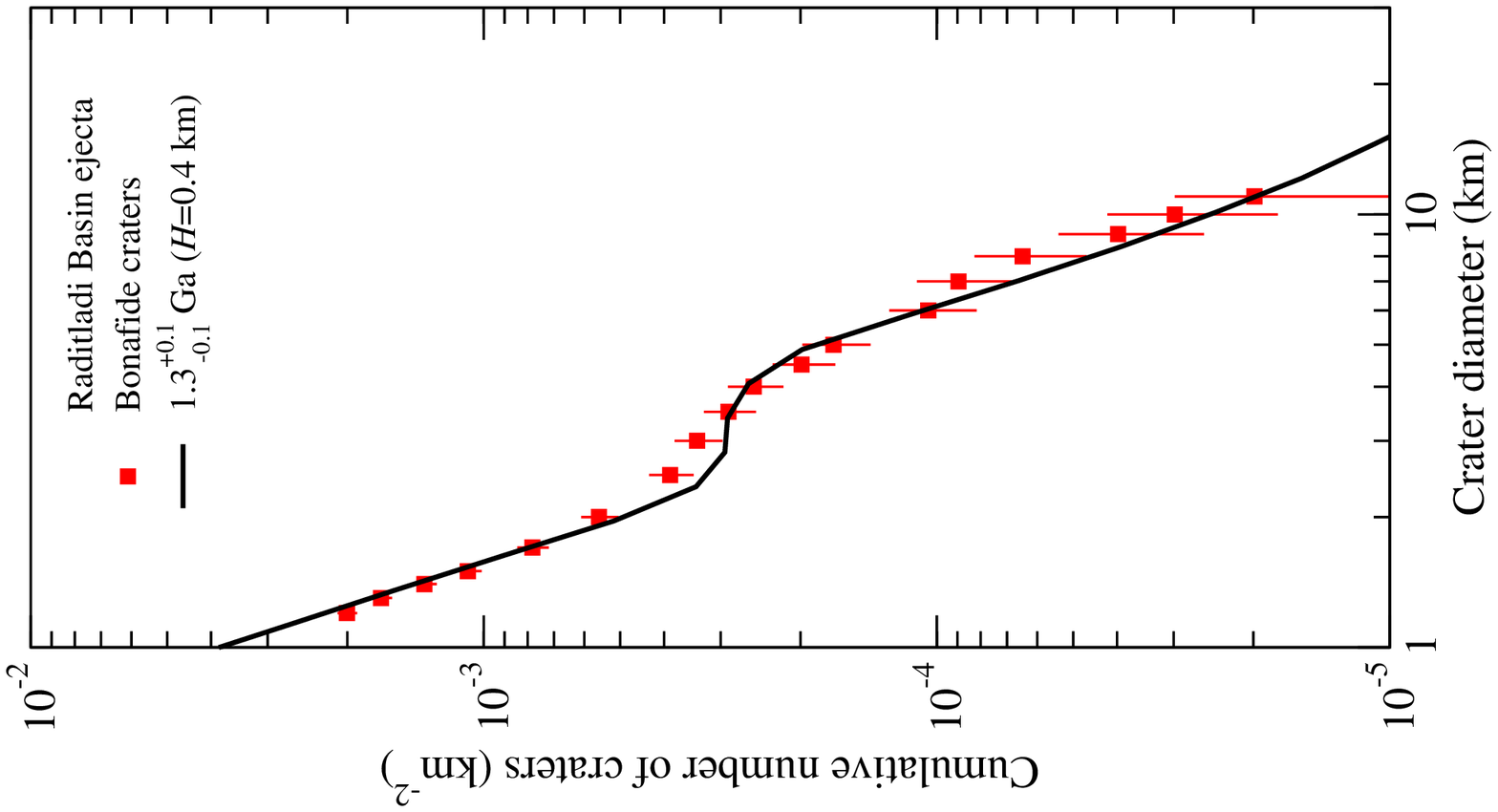}
\caption{Raditladi age assessment. MPF best fit of the bonafide
  crater SFDs for the inner plains and annular units (upper panels)
  and ejecta (lower panel).}
\label{rad_MPF}
\end{figure*}

\begin{figure*}[h]
\includegraphics[width=12cm]{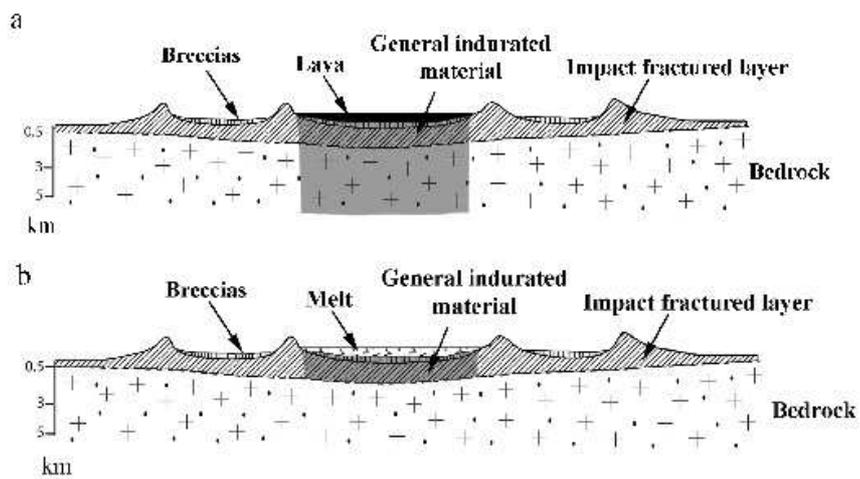}
\caption{ Possible geological sections of Raditladi basin hypothesized
  from the fit of MPF with the crater SFD of ejecta, annular materials
  and inner plains. Two scenarios can be argued to justify a very low
  value of $H$ for the inner plains a) lavas emplaced soon afterward
  the impact hardened the fractured and brecciated material within the
  basin or, b) impact melts completely hardened or replaced the impact
  breccias.}
\label{rad_scenario}
\end{figure*}

Raditladi contains an interior  peak-ring structure 125~km in diameter
and its  walls appear  to be degraded,  with terraces  most pronounced
within the north and west sides of the rim \citep{pro09}. A continuous
ejecta blanket with no visible  system of rays surrounds the basin and
extends up  to 225~km from  the basin rim (Fig.   \ref{rad_geo}).  The
floor is partially filled  with smooth, bright reddish plains material
that   clearly   embays   the   rim   and  the   central   peak   ring
\citep{ble09,pro09}. The  northern and  southern sectors of  the basin
floor  consist  of  dark,  relatively blue  hummocky  plains  material
confined between the rim and the peak ring. Troughs are found close to
the center  of the basin  arranged in a partially  concentric pattern,
$\sim$70~km  in  diameter,  and   are  interpreted  either  as  graben
resulting from post-impact  uplift of the basin floor,  or as circular
dikes     possibly     representing     fissural     feeding     vents
\citep{hea09,pro10}.  Floor material was subdivided into two different
units   following   \cite{pro09}:   smooth  and    hummocky
plains. Smooth plains may have a volcanic origin, as appears to be the
case  for  plains  in  the nearby  Caloris  basin  \citep{hea09,rob08},
however no  clear stratigraphic relation with the  hummocky plains has
been  found,  suggesting  that   all  the  different  terrains  within
Raditladi  basin may  be coeval  and  directly related  to the  impact
\citep{pro10}.  With respect to the  ejecta area, we have selected the
hummocky continuous ejecta blanket surrounding the basin.\\
We performed a crater count of the inner plains within the peak
  ring and the annular units enclosed between the basin rim and the
  peak ring.  Counts were also performed on the ejecta blankets.\\
Within the inner plains, numerous small graben-related pits up to 5~km
in diameter are identified; as for the case of Rachmaninoff, they are
most probably tectonically-originated features and/or volcanic vents
(fig.  \ref{rad_ex}, panels A and B).  Specifically, two peculiar pits
in the northern inner plains were interpreted as volcanic vents for
the dark material on the crater floor \citep{hea09}. Secondary craters
have been found associated to a 23-km crater within the inner plains
(fig.  \ref{rad_ex}, panels A and C). Some secondaries are also present
on the annular plains, whereas the ejecta blanket is characterized by
numerous self secondary craters, occurring mainly in clusters and
chains.  Figure \ref{rad_geo} shows the bonafide craters. The
statistics of all the identified features are reported in table
\ref{tab}, whereas the corresponding SFDs are shown in
fig. \ref{rad_SFD}.\\
The cumulative bonafide crater SFDs for different terrains of the
Raditladi basin are shown in fig.  \ref{rad_MPF}, along with the MPF
model ages. \\
The measured crater SFD on the ejecta blanket shows a flexure at about
$D=2-3$~km. The position of the flexure is well above the size of
craters that can no longer be distinguished because of the image
resolution, nevertheless the contribution of secondary craters at
these crater sizes might be important.  In the assumption that this
feature is due to the layering, the best fit achieved
for $H=0.4$~km gives a model age of $1.3\pm0.1$~Ga.  The best fit for
the annular units is achieved using $H=0.7$~km and the resulting model
age is $1.1\pm0.1$~Ga.  These values are consistent with both the
layering and model age inferred for the ejecta blankets. For this
reason the basin formation can be reliably fixed at around 1.1-1.3 Ga
in accordance with the age suggested by \cite{str08} on the basis of a
relative-chronology approach.  As for the Rachmaninoff basin, the poor
statistics and the limited range of diameters imply that the inner
plains SFD cannot be used to constrain $H$. We derived the model age
with both the same $H$ of the annular units ($H=0.7$~km), and a
negligible thickness ($H\sim0$), obtaining $2.2\pm0.3$~Ga and
$1.1\pm0.1$~Ga, respectively.  The former model age leads to a paradox
given that the annular units are certainly coeval with the basin
formation and, consequently, must only be older or of the same age as
the smooth plains.  Hence, the most reliable result for the inner
plains is to consider a solid material yielding a crater retention age
of about 1.1~Ga. The solid material could be due to an emplacement of lavas soon
afterward the impact leading to a complete hardening of the fractured
and brecciated material within the basin (fig. \ref{rad_scenario}a). 
This interpretation is
consistent with the presence of volcano-tectonic features within the
basin but may conflict with both the absence of distinctive color
variations of the inner plains with respect to the surrounding areas
and their unclear stratigraphic relationship with the annular
units. Alternatively, a great amount of impact melts able to
completely harden the impact breccias may explain the derived crater
retention age (fig. \ref{rad_scenario}b).

\section{Discussion and conclusions}

\begin{figure*}[h]
\includegraphics[width=12cm,angle=-90]{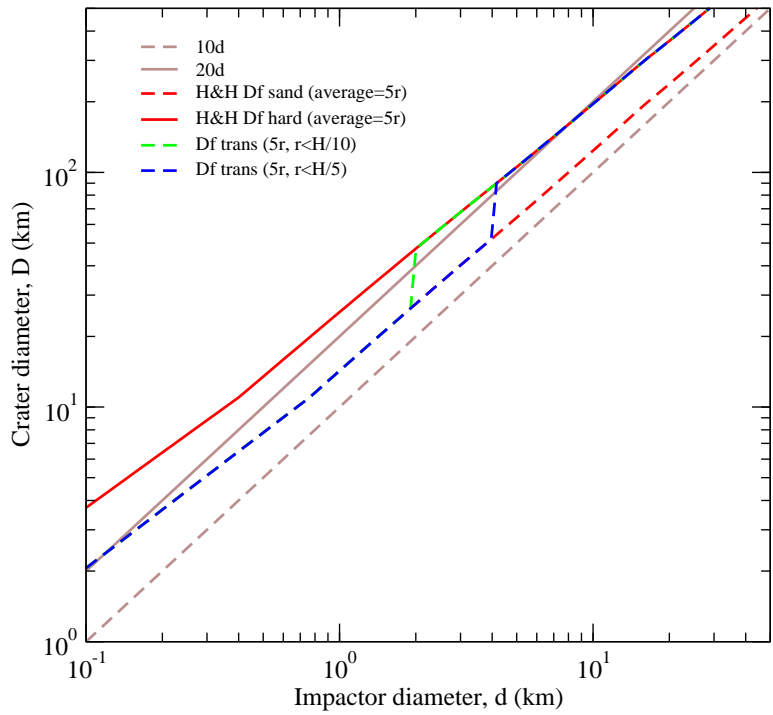}
\caption{Crater scaling laws from Mercury. The plot reports the crater
  scaling laws for both cohesive material (dashed red line) and
  hard-rock material (solid red line). Two examples of the scaling law
  corresponding to the layered target are shown. They correspond to a
  transition in the crater scaling law for impactor radius $r=H/10$
  (green dashed line) and $r=H/5$ (blue dashed line).  These values
  correspond to the predicted uncertenties in the modeling of the
  transition (see text), and in the rest of the work we adopt an
  intermediate value of $r=H/7.5$. The depth of the cohesive layer $H$
  has been set to 10~km and density and strength are averaged to a
  depth of $5r$.  In all cases, the curves are derived for the average
  impact velocity of 42~km/s, and an impact angle of $\pi/4$. The
  estimated impactor sizes that formed Rachmaninoff and Raditladi
  basins are also shown.}
\label{SL}
\end{figure*}

 MPF crater chronology has been applied to date the Rachmaninoff
  and Raditaldi basins. Age assesment has been performed taking into
  account target rheological layering and using the present NEO
  population as the prime source of impactors.\\
Our results demonstrate that the volcanic activity within the
Rachmaninoff basin interior significantly post-dates the formation of
the basin.  The basin itself probably formed about 3.6~Ga ago, whereas
the volcanic inner plains may have formed less than 1~Ga ago.
Therefore, Mercury had a prolonged volcanic activity, which possibly
persisted even longer than on the Moon, where the youngest detected
nearside flows \citep[on Oceanus Procellarum;][]{hie01} are about
1.1~Ga old.  On the other hand, the Raditladi basin has an estimated
model age of about 1~Ga and no firm indication that the inner plains
formed more recently than the basin itself.  Hence, these plains may be
due to either huge volumes of impact melts or lavas emplaced soon
afterward the basin formation.  In the latter case, which is not
clearly supported by the stratigraphic observations, volcanism might
have been triggered by the impact itself.\\
This work also shows the role of target properties in deriving the age
of a surface. Where such properties are neglected, as in traditional
chronologies \citep[e.g.][]{neu94}, the crater production function may
be unable to accurately reproduce the observed crater SFD and/or to
provide a consistent age for nearby terrains.  The following examples
serve to illustrate this point: the Rachmaninoff ejecta bonafide
crater SFD shows an S-shaped feature that, according to our best
knowledge, cannot be ascribed to processes other than a layered
target; the Raditladi inner plains have an higher density of craters
than the annular units, implying a paradoxically older age for the
interior plains if the inner and outer plains had the same material
properties.\\
The derived ranges of ages for Raditladi basin imply that its
formation occurred long after the late heavy bombardment
($\sim3.8$~Ga), at a time when the primary source of impactors was a
NEO-like population.  This conclusion also likely applies to
Rachmaninoff basin, even if it cannot be excluded that it was formed
during or prior the late heavy bombardment.  The NEOs average impact
velocity on Mercury is about 42~km/s \cite{mar05}.
Considering a most probable impact angle of $\pi/4$, the projectiles
responsible for Rachmaninoff and Raditladi formation should have had
diameters in the range 14-16~km (see fig. \ref{SL}).  In the present
NEO population, bodies are quickly replenished -in time scales of tens of
Myrs- mainly from the main belt via slow orbital migration into major
resonances.  Such a migration, due mainly to Yarkowsky effect, is size
dependent and is negligible for objects larger than $\sim10$~km.
Therefore larger objects, such as those required for the formation of
the Raditladi and Rachmaninoff basins, are mainly produced by
dynamical chaos loss \citep{min10}. Those simulations show that the
rate of large impactors decreased by a factor of 3 over the last 3~Ga.
Another source of large impactors is the sporadic direct injection
into strong resonances due to collisions \citep{zap97}. The present
NEOs' average impact probability with Mercury in the size range of
14-16~km, is of about one impact every 3.3~Ga, in agreement with the
proposed timescales of the formation of Rachmaninoff and Raditladi.\\

\newpage

{\bf Acknowledges}

The authors wish to thank P.~Michel for helpful discussions on the
cratering processes on a layered target. We also wish to thank
A.~Morbidelli for discussions regarding the NEO population. Finally,
we thank the referees (C. Chapman and an anonymous one) for providing
very interesting comments, that helped to improve our work.


\newpage

\end{document}